\documentclass[aps,prb,preprint,showpacs,amsmath,amssymb]{revtex4-1}

\usepackage{graphicx}
\usepackage{bm}
\usepackage{datetime}

\newcommand{\arm}{{\rm a}}

\newcommand{\qanh}{\left(Q^{\mathrm{ph}}_{m,\alpha}\right)^{-1}}

\begin{document}

\title{Phonon-mediated damping of mechanical vibrations in a finite atomic chain coupled to an outer environment}

\author{Ze'ev Lindenfeld, Eli Eisenberg, and  Ron Lifshitz}
\affiliation{Raymond and Beverly Sackler School of Physics and Astronomy,
Tel Aviv University, Tel Aviv 69978, Israel}

\date{September 23, 2013}
\begin{abstract}

We study phonon-mediated damping of mechanical vibrations in a finite quantum-mechanical atomic-chain model. Our study is motivated by the quest to understand the quality factors ($Q$) of nanomechanical resonators and nanoelectromechanical systems (NEMS), as well as actual experiments with suspended atomic chains and molecular junctions. We consider a finite atomic chain which is coupled to a zero-temperature outer environment, modeled as two additional semi-infinite chains, thus inducing ``clamping-losses''. Weak coupling to the outer environment ensures that the clamping losses are small, and that the initially discrete nature of the phonon spectrum is approximately maintained. We then consider a phonon damping process known as ``Landau-Rumer damping'', where phonons in the excited mode of vibration decay into other modes through anharmonic phonon-phonon interaction. The approximately discrete nature of the phonon spectrum leads to sharp nonmonotonic changes in $Q$ as parameters are varied, and to the appearance of resonances in the damping. The latter correspond to the existence of decay processes where the participating phonons approximately conserve energy. We explore means to control the damping by changing either the number of atoms in the chains or the ratio between the longitudinal and transverse speeds of sound, thereby suggesting future experiments to observe this resonance-like behavior.

\end{abstract}

\maketitle
\section{Introduction}\label{sec:introduction pp}

Understanding of the factors that determine $Q$ of nanomechanical resonators and of NEMS remains central for the development of their applications. These applications include low phase-noise oscillators;\cite{Kenig12} highly sensitive mass,\cite{ekinci04,*Yang06,*Hanay12,*ilic04,*Jensen08,*lassagne08} spin,\cite{rugar} and charge detectors;\cite{cleland98} and ultrasensitive thermometers \cite{roukes99} and displacement sensors.\ \cite{cleland,*knobel,*ekinci02,*truitt07} The ability to manufacture high-$Q$ nanoresonators is also important for basic research in the mesoscopic physics of phonons,\cite{Schwab_974} and the general study of the behavior of mechanical degrees of freedom at the interface between the classical and the quantum worlds.\cite{Schwab_36,*lahaye04,*naik06,oconnell2010,katz,katz08} A variety of different mechanisms---such as internal friction due to bulk or surface defects,\cite{Mihailovich95,*olkhovets,*carr2,*evoy2,*liu,*mohanty,*zolfagharkhani,*seoanez,*remus,*chu07,*unterreithmeier} damping due to phonon-phonon interaction,\cite{lifshitzTED,*houston,*sudipto,*kiselev,lifshitzPhonon} clamping losses,\cite{lifshitz2,*photiadis1,*geller1,*geller2,*geller3,*schmid08,*wilson-rae,*cole} and electron-phonon interaction \cite{lindenfeld2}---may contribute to the dissipation of energy in mechanical resonators and thus impose limits on their $Q$. The dissipated energy is transferred from a particular mode of the resonator, which is driven externally, to energy reservoirs formed by all other degrees of freedom of the system. Here, we focus our attention on damping due to phonon-phonon interaction arising from energy transfer between the driven mode and the rest of the modes of the nanoresonator.

The damping of sound waves due to phonon-phonon interaction in bulk insulators was extensively studied since the 1930's.\ \cite{landauRumer,akhiezer,woodruff} Phonon-phonon interaction also plays a central role in determining the lifetime of bulk phonons, as it is expressed in neutron scattering experiments.\ \cite{kokkedee2,maradudin2} One can consider several regimes of dissipation that are determined by the product $\omega\tau$, where $\omega$ corresponds to the frequency of the excited vibration while $\tau$ corresponds to the typical lifetime of the phonons involved.\ \cite{landauRumer,zimanPandE,reissland}

In the case where the lifetime of the phonons is so short that $\omega\tau\ll1$ (the thermoelastic regime) the phonons can be assumed to set a continuous uniform temperature field in the resonator which is disturbed by the excited vibration.\ \cite{lifshitzPhonon} The resulting heat flows, caused by the induced temperature gradients, lead to energy dissipation from the excited vibration. In the context of nanomechanical systems, thermoelastic damping was studied in micro- and nanomechanical resonators,\cite{lifshitzTED,*houston,*photiadis3,*sudipto} as well as in nanotubes.\ \cite{hajnayeb,*hoseinzadeh,*sun-bae} If $\omega\tau<1$ then the system operates within the ``Akhiezer regime''.\ \cite{akhiezer,lifshitzPhonon,tabrizian} Within this regime, which was also investigated for nanomechanical resonators,\cite{goldfarb,kiselev,kunal} the externally excited vibration can be regarded as a macroscopic strain field that modulates locally the lattice constant of the material and, as a result, the frequency of the thermal phonons of the resonator, thus driving them out of equilibrium. The process of restoring the local thermal equilibrium to the phonon population results in the dissipation of energy from the excited vibration. Lastly, in this study we concentrate on the ``Landau-Rumer regime'' where $\omega\tau>1$.\ \cite{landauRumer,lifshitzPhonon,tabrizian}

In the Landau-Rumer regime an externally excited vibration of a given frequency in the nanomechanical resonator can be considered as an addition of phonons to a given mode or modes of vibration of the resonator. The condition $\omega\tau>1$ ensures that these added phonons are damped by interaction with other individual well-defined phonons of the resonator that are in thermal equilibrium.\ \cite{zimanPandE,reissland} The interaction occurs due to the anharmonic nature of the resonator, which can be  treated within the framework of perturbation theory.\ \cite{landauRumer,zimanPandE,reissland,ashcroft} Thus, the lifetime of the phonons cannot be too short, which sets a limit on the damping caused by the different dissipation mechanisms. This implies that in order to operate within the Landua-Rumer regime one needs to consider a resonator that is only weakly coupled to the outer environment, that is sufficiently clean, and that operates at a low temperature. The fact that ballistic transport of phonons was observed in nanomechanical structures at sufficiently low temperatures \cite{Schwab_974} indicates that the Landau-Rumer regime could be reached in nanomechanical resonators. There are also indications that the lifetime of certain vibrational modes in ultrathin silicon membranes is determined by Landau-Rumer damping even at room temperature.\ \cite{cuffe}

For nanomechanical resonators, the Landua-Rumer regime was studied theoretically---mainly in carbon nanotubes \cite{xiao2003,*hepplestone2006,martino,*ong}, nanowires,\cite{khitun,*lu1} and atomic chains.\ \cite{goldfarb} However, in almost all of these studies the vibrational modes of the resonator were assumed to be bulk-like, or at least with a continuous spectrum, because one or more of the dimensions of the system were taken to be effectively infinite. We consider a confined mesoscopic resonator, where the vibrational modes are ideally discrete, which operates within the Landau-Rumer regime, and study phonon-phonon damping in the presence of mechanical coupling to an outer environment with a continuum of vibrational modes.

We examine phonon-mediated damping within a simple yet concrete physical model, similar to that of Goldfarb,\cite{goldfarb} of a finite atomic chain, coupled to an outer environment. Coupling to the outer environment leads to clamping-losses,\cite{lifshitz2,*photiadis1,*geller1,*geller2,*geller3,*schmid08,*wilson-rae,*cole} namely to radiation of mechanical energy to the outer environment, which is unavoidable in many realistic systems. Furthermore, in the absence of coupling to the outer environment, the spectrum of the vibrational modes of the finite atomic chain (or any other finite system) remains discrete, and the density of states (DOS) of the phonons consists of a sum of $\delta$-function peaks. The anharmonic interaction between such infinitely sharp phononic states does not lead to dissipation, but rather to oscillations between the various states of the finite chain. Dissipation can occur only because of an interaction with states that form a continuum. In a previous study \cite{lindenfeld2} we considered the contribution of electron-phonon interaction to the damping of phonons in metallic nanomechanical beams. There, the continuum of states was formed by the electronic degrees of freedom. Here, since we consider phonon-phonon interaction within the finite chain, the continuum of states needs to come from the phonons themselves. The interaction with the outer environment broadens the phonon levels and their spectral functions and changes the discrete DOS into a continuous one. Although we concentrate on the study of a model for an atomic chain, our qualitative results should also apply to other systems, such as nanoparticles and short nanowires or nanotubes, in which confinement of phonons plays an important role.

Atomic chains or molecular wires, composed of several atoms ordered in a linear chain, were manufactured in various configurations using several types of materials. Metallic atomic chains suspended between two electrodes were manufactured from gold \cite{ohnishi,yanson} and silver \cite{rodrigues} atoms. Mixing between different atomic species within the same linear chain was also accomplished.\ \cite{bettini} In addition, artificially-made as well as self-assembled metallic atomic chains formed on a surface were also fabricated using gold,\cite{*nilius1,*wangJ} palladium,\cite{nilius2} platinum,\cite{gurlu} and bismuth,\cite{miki} as well as semiconducting materials such as silicon and germanium (see the study of Soukiassian \cite{soukiassian} and references therein).

Phonons in atomic chains and molecular junctions were studied mainly in the context of their influence on electronic transport. The fingerprints of excited vibrational modes in the electronic conductance of atomic chains were detected experimentally,\cite{agrait1,*agrait2} and studied theoretically.\ \cite{frederiksen,*kushmerick,*viljas,*paulsson,*vega,galperin} Thermal transport through atomic chains was also studied, mainly by employing noninteracting phonon models within a single-particle scattering Landauer-like approach or classical molecular dynamics (see the study of Dubi and Di Ventra \cite{dubi} and references therein). Only a limited number of studies examined the role of quantum phonon-phonon scattering on thermal conductance through atomic chains.\ \cite{mingo,WangJianSheng,li014308} These studies involve rather complex models, mainly due to the detailed description of the thermal baths and their coupling to the atomic chain. Although we use a simple and less realistic model, it enables us to easily incorporate the effects of the outer environment, and examine effects that arise from the finite size of the system.

The effect of phonon-phonon interaction on the lifetime of phonons in infinite 1d chains was addressed in the 1960's.\ \cite{maradudin,conway,meier,pathak} In these studies, the contribution of only three-phonon processes was considered within the framework of the lowest order perturbation theory. Here we follow the same approach when we calculate the imaginary part of the self-energy of the phonons of the finite chain that results from the phonon-phonon interaction, and from it the damping rate of these phonons. The higher-order phonon processes, as well as higher-order terms in the perturbation expansion of the three-phonon processes, can be of importance. For example, in the bulk the simplest three-phonon events can be negligible due to restrictions imposed by crystal-momentum and energy conservation laws. Therefore, the lifetime of the phonons is determined by four-phonon or even five-phonon processes, as well as by higher order terms in the perturbation theory of three-phonon processes.\ \cite{ashcroft,zimanPandE}

We restrict ourselves to zero temperature, which results in simpler analytical and numerical expressions. Higher temperature leads to higher damping rates due to phonon-phonon interaction. However, it also smears the effects of the discrete nature of the phonons. Furthermore, the condition $\omega\tau>1$ might be violated at high temperatures, and the resonator would not operate within the Landau-Rumer regime. An additional simplification is introduced by considering relatively short chains containing a few tens of atoms at most. We expect the effect of the discrete nature of the DOS to be more pronounced in such short chains. Furthermore, suspended atomic chains that are realized experimentally typically contain only a few atoms.\ \cite{ohnishi,yanson,rodrigues,bettini} We consider a chain and an environment made of a semiconducting (at zero temperature, insulating) material, where there are no electron-phonon contributions, in order to focus on the effects of phonon-mediated damping alone. We obtain numerical results by using the relevant average physical properties of bulk silicon. Using these parameters suffices for our purposes since we only want to study the qualitative features of the damping, and do not aim for a realistic description of an atomic chain.

The structure of the article is as follows. In section \ref{sec:A finite harmonic chain coupled to an outer environment} we describe our model of a finite harmonic chain coupled to an outer harmonic environment, obtain the Green function and spectral function of the broadened phonons of the finite chain, and present the qualitative features of the clamping-losses. In section \ref{sec:Anharmonic interaction within the finite chain}, we introduce anharmonic phonon-phonon interaction between the broadened phonons of the finite chain, calculate the imaginary part of the self-energy of the phonons of the finite chain that results from the phonon-phonon interaction, and present numerical results that demonstrate the main features of the additional damping. We summarize our results and conclusions in section \ref{sec:Conclusions pp}.

\section{A finite harmonic chain coupled to an outer environment}\label{sec:A finite harmonic chain coupled to an outer environment}

\subsection{Eigenmodes of an infinite chain with two heavy masses}\label{subsec:Eigenmodes of an infinite chain with two heavy masses}

We consider an infinite chain placed along the $x$ axis with atoms of mass $m_{a}$ and lattice spacing $d$. We fix the origin ($n=0$) at the middle site of the finite chain.  The atoms at the sites $n=\pm n_{0}$ are replaced with heavier atoms with mass $M_{a}$. The finite atomic chain is therefore located between $|n|<n_{0}$, and the heavy masses act as clamps that couple the finite chain to the two semi-infinite chains that form the outer environment. We assume a nearest-neighbor interaction between the atoms, which is taken to be the same whether it acts between two identical atoms or between an atom and one of the clamps. The coupling between the finite chain and the two external semi-infinite chains can be tuned by changing the mass ratio $r=M_{\arm}/m_{\arm}$.

The entire infinite chain can vibrate in three directions---in the longitudinal $x$ direction, along the chain, and in the transverse $y$ and $z$-directions, i.e.\ perpendicular to the chain. The potential energy of the chain consists of a harmonic part $U_{2}$ and an anharmonic part $U_{3}$. The harmonic term is
\begin{equation}\label{eq:U2}
U_{2}=\frac{1}{2}\sum_{n,\alpha}\phi_{\alpha}\left(u_{\alpha}(n+1,t)-u_{\alpha}(n,t)\right)^{2},
\end{equation}
where $u_{\alpha}(n,t)$ is the displacement of the $n^{th}$-atom from its equilibrium position in the $\alpha=x,y,z$ direction, and $\phi_{\alpha}$ are the harmonic force constants. The kinetic energy term is
\begin{equation}\label{eq:kinetic inf}
T=\sum_{n\neq \pm n_{0},\alpha}\frac{1}{2}m_{\arm}\left(\partial_{t}u_{\alpha}(n,t)\right)^{2}+\frac{1}{2}M_{\arm}\left[\left(\partial_{t}u_{\alpha}\left(n_{0},t\right)\right)^{2}+
\left(\partial_{t}u_{\alpha}\left(-n_{0},t\right)\right)^{2}\right],
\end{equation}
and the harmonic part of the Hamiltonian of the entire infinite chain is
\begin{equation}\label{eq:H2 inf}
H_{2}=U_{2}+T.
\end{equation}
The harmonic force constants are calculated from bulk silicon longitudinal and transverse speeds of sound, $c_{l}$ and $c_{t}$ respectively, using the relation $c_{l,t}=d\sqrt{\phi_{l,t}/m_{\arm}}$, where $\phi_{l}=\phi_{x}$ and $\phi_{t}=\phi_{y}=\phi_{z}$. The eigenmodes and eigenfrequencies that result from the Hamiltonian \eqref{eq:H2 inf} are readily calculated (see, for example, the study by Fukuda \cite{fukuda} for the calculation of the eigenfrequencies in the general case where $r$ can be smaller than 1). For the sake of completeness, we give here the detailed results.

The Hamiltonian $H_{2}$ of Eq.\ \eqref{eq:H2 inf} yields equations of motion that possess eigenmode solutions of the form
\begin{equation}\label{psi time dep}
\Psi_{k,\alpha}(n,t)=\Psi_{k,\alpha}(n)e^{i\omega_{k,\alpha}t},
\end{equation}
where the modes $\Psi_{k,\alpha}(n)$ are composed of left and right moving plane waves. We choose the eigenmodes to be real functions of $n$, and express them using symmetric and antisymmetric functions $\Psi_{k,p,\alpha}(n)$, where the $p=s$ or $p=a$ denotes the parity of the eigenmode
\begin{equation}\label{eq: un infinite}
\Psi_{k,p,\alpha}(n)=\frac{B_{k,p}}{\sqrt{N}}
\begin{cases}
\left(\delta_{k,p}e^{ikdn}+\epsilon_{k,p}e^{-ikdn}\right), \; &n_{0}\leq n,\\
\left(e^{ikdn}\pm e^{-ikdn}\right), \; &-n_{0}\leq n\leq n_{0}, \\
\pm\left(\epsilon_{k,p}e^{ikdn}+\delta_{k,p}e^{-ikdn}\right), \; &n\leq -n_{0}.
\end{cases}
\end{equation}
The upper (lower) sign in Eq.\ \eqref{eq: un infinite} corresponds to the symmetric (antisymmetric) solution, and $N$ is the total number of atoms in the entire chain (including the two outer chains) which is taken to be a large number that tends to infinity.

By substituting either the symmetric or antisymmetric solutions in the equations of motion we obtain a dispersion relation that is identical to the one obtained for a homogenous infinite chain
\begin{equation}\label{eq:dispersion relation}
\omega_{k,\alpha}=\omega_{0,\alpha}\left|\sin{\frac{1}{2}kd}\right|,
\end{equation}
where $\omega_{0,\alpha}=2\sqrt{\frac{\phi_{\alpha}}{m_{\arm}}}$.

The continuity condition of the solution at $n=n_{0}$, together with the equation of motion at $n=n_{0}$, and the dispersion relation \eqref{eq:dispersion relation} determine the ratios
\begin{equation}\label{CdivA}
\delta_{k,p}=-i(r-1)\tan{\left(\frac{kd}{2}\right)}\left(\mp e^{-2in_{0}kd}-1\right)+1
\end{equation}
and
\begin{equation}
\epsilon_{k,p}=-i(r-1)\tan{\left(\frac{kd}{2}\right)}\left(e^{2in_{0}kd}\pm1\right)\pm1,\label{DdivA}
\end{equation}
where, again, the upper (lower) sign in Eqs.\ \eqref{CdivA} and \eqref{DdivA} corresponds to the symmetric (antisymmetric) solution. In the limit of $r=1$ Eqs.\ \eqref{CdivA} and \eqref{DdivA} lead to $\delta_{k,p}=1$ and $\epsilon_{k,p}=\pm 1$, as expected for a homogeneous infinite chain.

The quantized displacement field is given by
\begin{equation}\label{eq:quantization infinite chain}
\hat{u}_{\alpha}(n)=\sum_{k,p}\sqrt{\frac{\hbar}{2m_{\arm}\omega_{k,\alpha}}}\Psi_{k,p,\alpha}(n)A_{k,p,\alpha},
\end{equation}
where
\begin{equation}\label{eq:def operator A inf}
A_{k,p,\alpha}=a_{k,p,\alpha}+a_{k,p,\alpha}^{\dag},
\end{equation}
and $a_{k,p,\alpha}$ ($a_{k,p,\alpha}^{\dag}$) annihilates (creates) a phonon of the infinite chain with a wave number $k$, parity $p$, and polarization $\alpha$. In Eq.\ \eqref{eq:quantization infinite chain} the eigenmodes $\Psi_{k,p,\alpha}(n)$ are the dimensionless wave-functions of the phonons of the infinite chain, normalized according to
\begin{equation}\label{eq:norm discrete}
\sum_{n}\left|\Psi_{k,p,\alpha}(n)\right|^{2}=1.
\end{equation}
The normalization condition Eq.\ \eqref{eq:norm discrete} determines the normalized coefficient $B_{k,p}$
\begin{equation}\label{eq:A s and a}
B_{k,p}=\frac{1-(1+i)\delta(p,a)}{\sqrt{\left|\delta_{k,p}\right|^{2}+\left|\epsilon_{k,p}\right|^{2}}},
\end{equation}
where $\delta(p,a)$ is equal to 0 if $p=s$ and to 1 if $p=a$.

\subsection{Effect of clamping-losses}\label{subsec:The spectral function of the finite harmonic chain}

We want to examine the damping of vibrations confined to the finite chain by clamping-losses. In order to do so we consider the eigenmodes of a perfectly clamped finite chain, i.e.\ with $u\left(\pm n_{0},t\right)=0$. These modes are not eigenmodes of the full infinite-chain Hamiltonian $H_{2}$ defined in Eq.\ \eqref{eq:H2 inf}, and therefore their lifetime is finite. The finite lifetime is expressed as a broadening of the spectral function of the phonons of the finite chain.

The normalized eigenmodes of a finite chain of $2n_{0}-1$ atoms are
\begin{equation}\label{eq:finite modes}
\psi_{m,\alpha}(n)=
\begin{cases}
\frac{1}{\sqrt{n_{0}}}\cos{\left(q_{m}dn\right)}, & m=1,3,\ldots,2n_{0}-1 \\
\frac{1}{\sqrt{n_{0}}}\sin{\left(q_{m}dn\right)}, & m=2,4,\ldots,2n_{0}-2
\end{cases}
\end{equation}
where $q_{m}=m\pi/2n_{0}d$, and $\psi_{m,\alpha}(n)=0$ for $|n|\geq n_{0}$. The modes with odd $m$ are symmetric, while modes with even $m$ are antisymmetric. The quantized displacement field within the finite chain is
\begin{equation}\label{eq:quantization finite chain}
\hat{u}^{\mathrm{fin}}_{\alpha}(n)=\sum_{m}\sqrt{\frac{\hbar}{2m_{\arm}\omega_{m,\alpha}}}\psi_{m,\alpha}(n)A_{m,\alpha},
\end{equation}
where
\begin{equation}\label{eq:def A operator fin}
A_{m,\alpha}=a_{m,\alpha}+a^{\dag}_{m,\alpha}.
\end{equation}
The operator $a_{m,\alpha}$ ($a^{\dag}_{m,\alpha}$) annihilates (creates) a phonon in an eigenmode of the perfectly clamped finite chain. Similarly to the quantized field in Eq.\ \eqref{eq:quantization infinite chain}, the eigenmodes $\psi_{m,\alpha}(n)$ in Eq.\ \eqref{eq:quantization finite chain} are dimensionless wave-functions of the phonons of the finite chain, which are normalized according to the condition
\begin{equation}\label{eq:norm condition finite}
\sum_{n=-\left(n_{0}-1\right)}^{n_{0}-1}\left|\psi_{m,\alpha}(n)\right|^{2}=1.
\end{equation}

The Green function of the phonons of the finite chain at zero temperature, written in the Lehmann representation,\ \cite{mahan,fetter} is
\begin{equation}\label{eq:green phonon}
D_{m,\alpha}^{0}(\omega)=\sum_{k,\beta} \left|\left\langle 1_{k,p,\beta}\right|A_{m,\alpha}\left|0\right\rangle\right|^{2}\frac{1}{\hbar}\left(\frac{1}{\omega-\omega_{k,\alpha}+i\epsilon }-\frac{1}{\omega+\omega_{k,\alpha}-i\epsilon }\right),
\end{equation}
where $\epsilon$ is an infinitesimally small quantity that tends to zero at the end of the calculation, $\left|0\right\rangle$ is the ground state (zero-phonon state) of the finite chain, and $\left| 1_{k,p,\beta}\right\rangle$ is a state of the entire infinite chain with a single excited phonon of wave number $k$, parity $p$, and polarization $\beta$. The expression for the Green function of Eq.\ \eqref{eq:green phonon} is obtained directly from the definition of the zero-temperature Green function $D_{m,\alpha}^{0}(\omega)=-i\left\langle0\right|TA_{m,\alpha}(t)A_{m,\alpha}(t')\left|0\right\rangle$, where $T$ is the time-ordering operator.\ \cite{mahan} The retarded Green function $D^{0,\mathrm{ret}}_{m,\alpha}(\omega)$ is obtained by replacing the $-i\epsilon$ in the second term in the brackets in Eq.\ \eqref{eq:green phonon} with $+i\epsilon$. The matrix elements of the operator $A_{m,\alpha}$ that appear in Eq.\ \eqref{eq:green phonon} are given by the overlap within the finite chain between the wave function of the mode of the finite chain $\left\langle n\right|A_{m,\alpha}\left|0\right\rangle$ and the wave function of the mode of the entire infinite chain $\left\langle n|1_{k,p,\beta}\right\rangle$
\begin{equation}\label{eq:matrix elements}
\left\langle 1_{k,p,\beta}\right|A_{m,\alpha}\left|0\right\rangle=\sum_{n=-\left(n_{0}-1\right)}^{n_{0}-1}\psi_{m,\alpha}\left(n\right)\Psi_{k,p,\beta}(n).
\end{equation}
The matrix elements vanish unless the excited phonon of the finite chain has the same polarization and parity as the phonon of the infinite chain. Thus $\alpha=\beta$, while $p=s$ if the mode number $m$ is odd and $p=a$ if $m$ is even. From now on we drop the parity index $p$ of the modes of the infinite chain since it is determined by the mode number $m$ of the mode of the finite chain.

The phonon spectral function $F_{m,\alpha}\left(\omega\right)$ is defined as \cite{mahan}
\begin{equation}\label{eq:spectral def}
F_{m,\alpha}\left(\omega\right)\equiv-2\mathrm{Im}\left[\lim_{\epsilon\rightarrow0}D^{0,\mathrm{ret}}_{m,\alpha}(\omega)\right].
\end{equation}
Taking the limit $\epsilon\rightarrow0$ of Eq.\ \eqref{eq:spectral def} and calculating the imaginary part of the result we obtain
\begin{equation}\label{eq:spectral1}
F_{m,\alpha}\left(\omega\right)=\frac{2\pi}{\hbar}\sum_{k} \left|\left\langle 1_{k,\alpha}\right|A_{m,\alpha}\left|0\right\rangle\right|^{2} \left[\delta\left(\omega-\omega_{k,\alpha}\right)-\delta\left(\omega+\omega_{k,\alpha}\right)\right].
\end{equation}
We use a dimensionless frequency $\Omega=\omega/\omega_{0,\alpha}$, and give our results in terms of the dimensionless spectral function $f_{m}\left(\Omega\right)=F_{m,\alpha}(\omega/\omega_{0,\alpha})\times\hbar\omega_{0,\alpha}$.

We insert the expressions for $\psi_{m,\alpha}\left(n\right)$, given in Eq.\ \eqref{eq:finite modes}, and for $\Psi_{k,p,\alpha}(n)$ (between the clamps $|n|<n_{0}$), given in Eq.\ \eqref{eq: un infinite}, into the expression \eqref{eq:matrix elements} for the matrix elements. We then change the sum in Eq.\ \eqref{eq:spectral1} into an integral, use the dispersion relation \eqref{eq:dispersion relation} to change the $\delta$ functions in Eq.\ \eqref{eq:spectral1} into $\delta$ function of the wave number $k$, and obtain
\begin{equation}\label{eq:spectral2}
f_{m}\left(\Omega\right)=
\begin{cases}
\mathrm{sign}(\Omega)\frac{4\left|\left\langle 1_{\kappa,\alpha}\right|A_{m,\alpha}\left|0\right\rangle\right|^{2}}{\sqrt{1-\Omega^{2}}}, &\left|\Omega\right|<1\\
0, &\left|\Omega\right|>1,
\end{cases}
\end{equation}
where $\kappa=2\sin^{-1}\Omega$. In Eq.\ \eqref{eq:spectral2} and from now on whenever a matrix element of the form $\left\langle 1_{\kappa,\alpha}\right|A_{m,\alpha}\left|0\right\rangle$ appears, it is multiplied by $\sqrt{N}$ which cancels the normalization factor $1/\sqrt{N}$ that normalizes the wave functions of the phonons of the entire chain in Eq.\ \eqref{eq: un infinite}. This normalization factor cancels when changing a sum over $k$ into an integral. The spectral function is therefore
\begin{equation}
f_{m}\left(\Omega\right)=\mathrm{sign}(\Omega)\frac{4}{\sqrt{1-\Omega^{2}}}\frac{\left|B_{\kappa}\right|^{2}}{n_{0}}
g_{m}\left(\kappa\right),\label{eq:spectral fun final}
\end{equation}
where
\begin{equation}\label{eq:g}
g_{m}\left(\kappa\right)=
\left[\frac{\sin{\left(2n_{0}\Theta^{-}_{m}\right)}\sin{\left(q_{m}d\right)}}{2\sin{\left(\Theta^{-}_{m}\right)}\sin{\left(\Theta^{+}_{m}\right)}}\right]^{2},
\end{equation}
$\Theta^{\pm}_{m}=\left(q_{m}d\pm \kappa\right)/2$, and $B_{\kappa}$ is the coefficient given in Eq.\ \eqref{eq:A s and a} with the parity index $p$ dropped. The dimensionless spectral function \eqref{eq:spectral fun final} is independent of the material parameters and polarization, and is fully determined by the mode number $m$ and the ratio $r$.

For chains and mass ratios $r$ considered in this study, most of the broadened spectral functions can be approximated rather well by a Lorentzian normalized to $2\pi$
\begin{equation}\label{eq:lorentizan approx}
f_{m}\left(\Omega\right)\simeq
\frac{2\Gamma^{\mathrm{c}}_{m}}
{\left(\Omega-\Omega_{m}^{*}\right)^{2}+\left(\Gamma^{\mathrm{c}}_{m}\right)^{2}}.
\end{equation}
The center $\Omega_{m}^{*}$ of the Lorentzian coincides with the center of the main peak of $f_{m}\left(\Omega\right)$, and $\Gamma^{\mathrm{c}}_{m}$ is half the width at half the maximum of the main peak. The dimensionless frequency $\Omega_{m}^{*}$ corresponds to the shifted frequency of the eigenmode, while $\left(2\Gamma^{\mathrm{c}}_{m}\right)^{-1}$ gives the finite lifetime of the mode owing to clamping-losses. We use the quality factor
\begin{equation}\label{eq:Q-1 clamping}
Q^{\mathrm{c}}_{m}=\frac{\Omega_{m}}{2\Gamma^{\mathrm{c}}_{m}},
\end{equation}
as a measure of the damping of the vibrational modes of the finite chain due to clamping-losses. The spectral functions of the low-frequency phonons deviate from the Lorentzian approximation, a deviation that becomes more severe as the frequency of the mode is lowered and the ratio $r$ is decreased. Note that our calculation is not restricted to the weak coupling regime, where the Lorentzian approximation is applicable, and the resulting Green functions and spectral functions can be used for any value of $r$.

\subsection{Clamping-losses -- results and discussion }\label{subec:Results harmonic}

The broadening of the discrete modes of the finite chain is necessary in order to consider the damping due to phonon-phonon interaction within the finite chain. Thus the entire analysis of the clamping-losses can be considered as a preliminary stage to the study of the damping induced by the phonon-phonon interaction, and should not be taken as a serious attempt to evaluate the clamping-losses for a realistic system. Nevertheless, although we use a simple model, the features of the resulting spectral function are not completely trivial. Therefore, in this section we describe in some detail these features and their origin.

\begin{figure} [!h]
\begin{center}
 \includegraphics{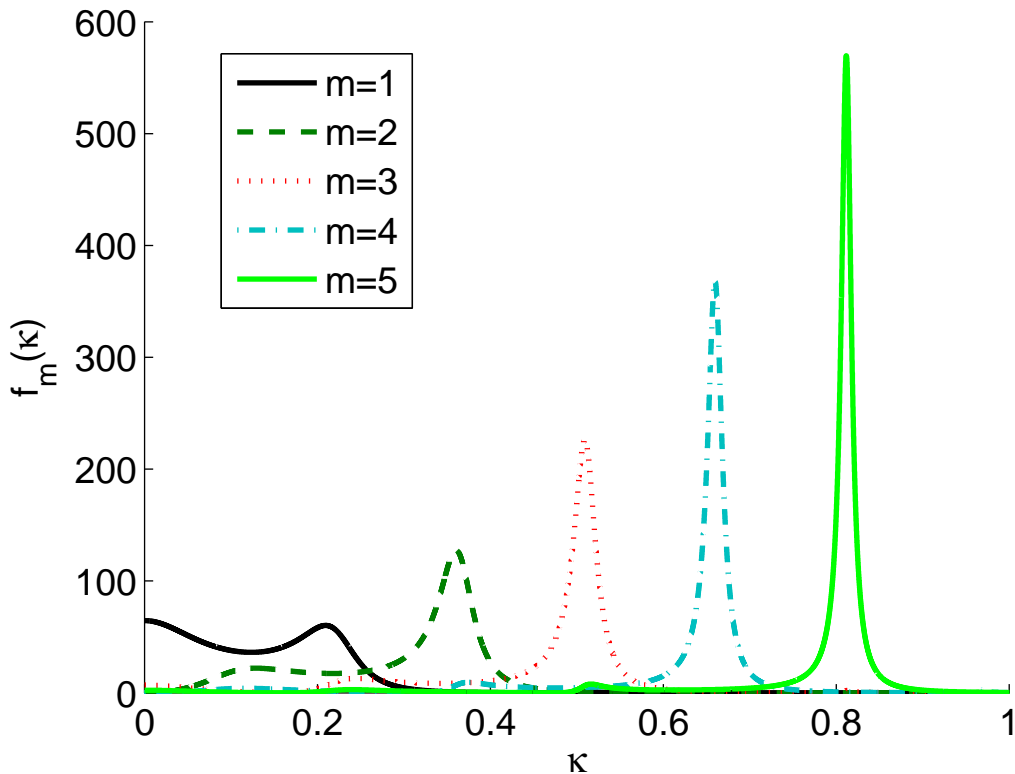}
 \caption{\label{fig1pp}
 (color online) The dimensionless spectral functions of the 5 lowest vibrational modes of a finite chain with $n_{0}=10$ (19 atoms) and $r=5$.}
 \end{center}
\end{figure}

\begin{figure} [!h]
 \begin{center}
 \includegraphics{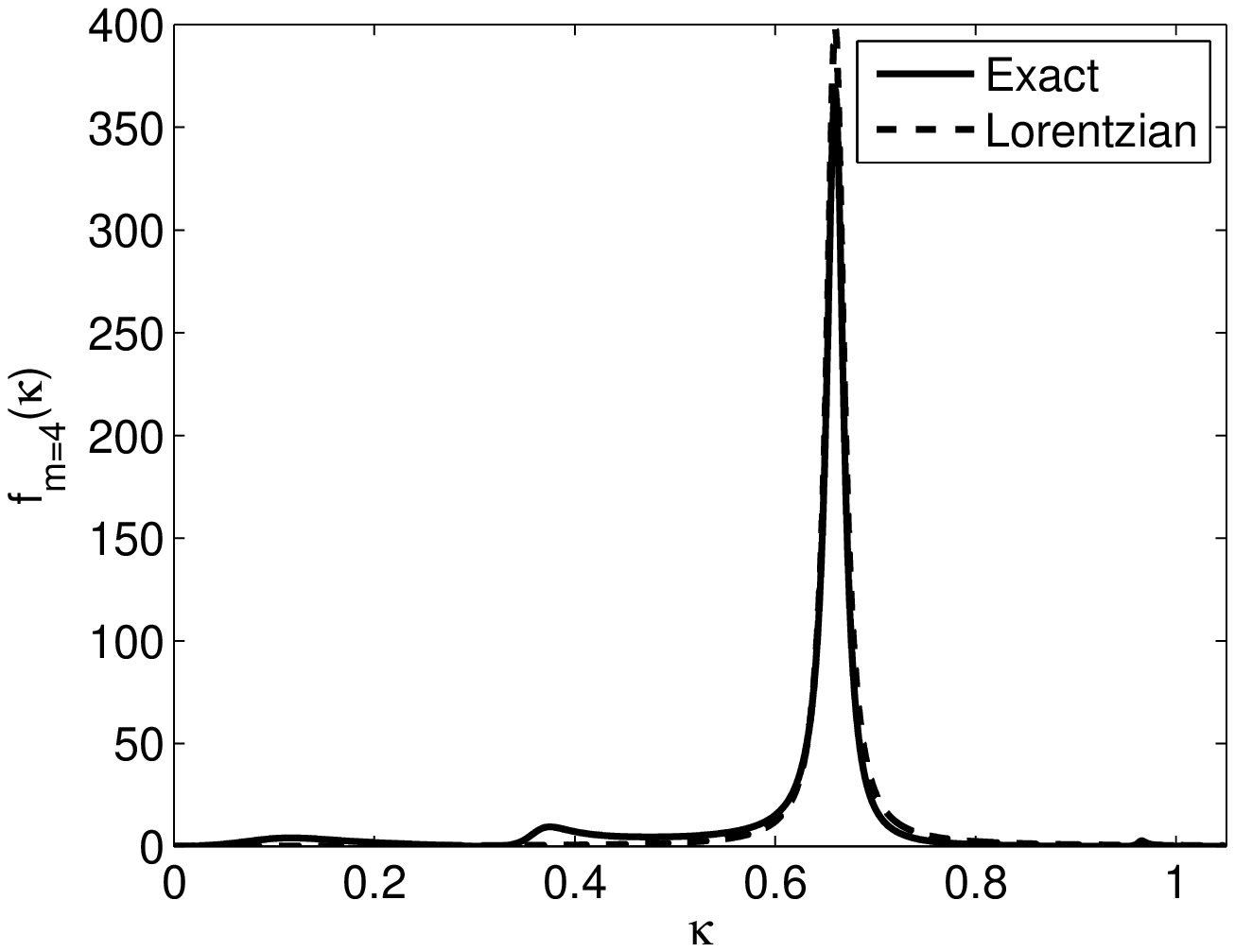}
 \caption{\label{fig2pp}
 A comparison between the exact dimensionless spectral function of the mode $m=4$ of the same chain that is considered in Fig.\ \ref{fig1pp} and its Lorentizan approximation.}
 \end{center}
\end{figure}

We consider a chain with $n_{0}=10$ and $r=5$, for which we plot in Fig.\ \ref{fig1pp} the dimensionless spectral function of the lowest five, longitudinal or transverse, modes. The results are plotted as a function of $\kappa$ rather than $\Omega=\sin{\frac{\kappa}{2}}$. Fig.\ \ref{fig1pp} demonstrates the improvement in the Lorentzian approximation for the spectral function with the increase in the mode number. A more detailed comparison of the Lorentizan approximation to the exact spectral function is shown in Fig.\ \ref{fig2pp} for the mode $m=4$.

\begin{figure} [!h]
 \begin{center}
 \includegraphics{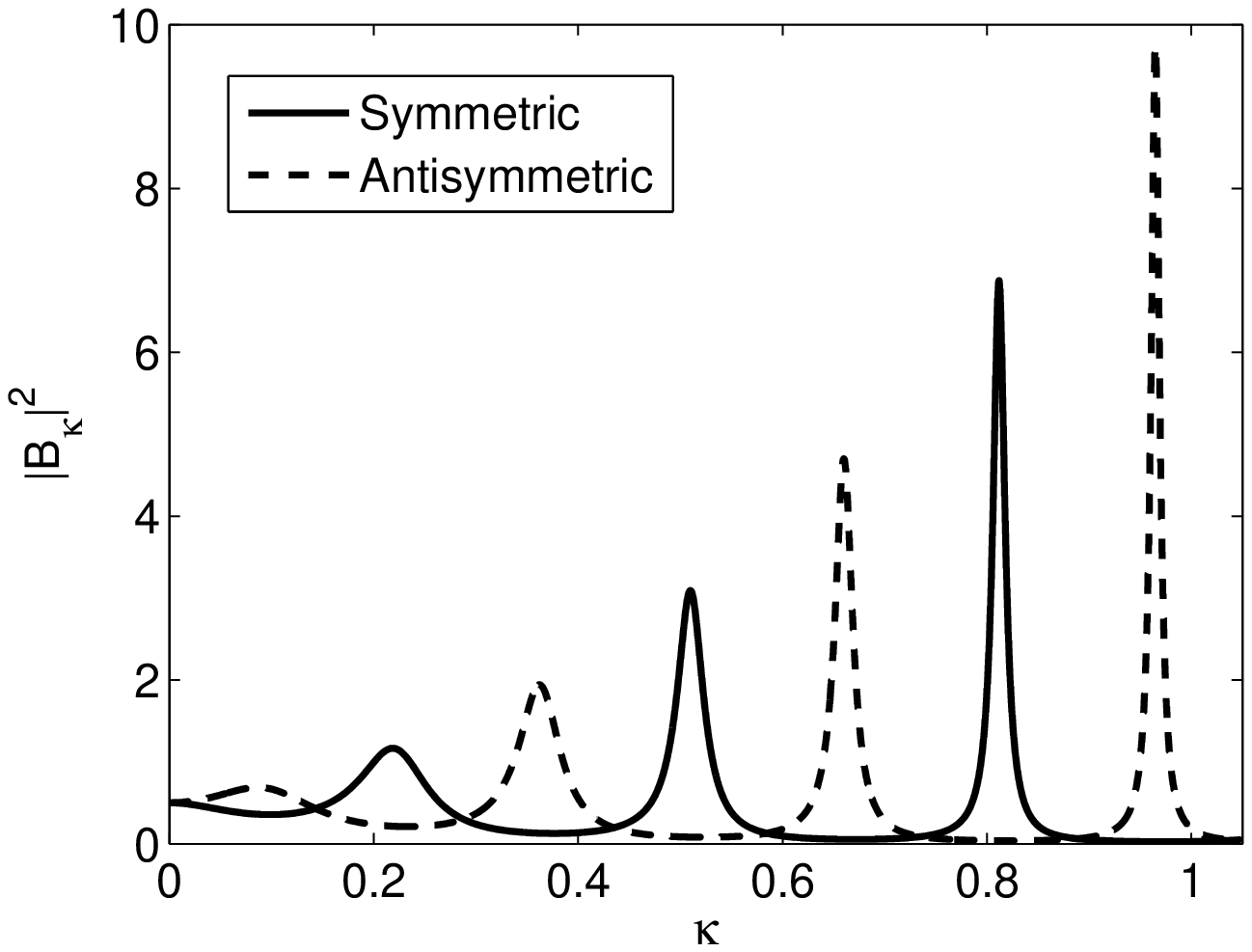}
 \caption{\label{figApp}
 The three lowest resonances of the squared amplitude $\left|B_{\kappa}\right|^{2}$ for both symmetric and antisymmetric modes.}
 \end{center}
\end{figure}

The spectral function \eqref{eq:spectral fun final} is controlled mainly by two factors---the squared coefficient $\left|B_{\kappa}\right|^{2}$ and the function $g_{m}\left(\kappa\right)$. The behavior of $\left|B_{\kappa}\right|^{2}$ is largely determined by the fact that unless $\kappa$ approximately matches one of the wave numbers of the finite chain the vibration of the atoms within the finite chain is small. However, around resonance wave numbers, which are approximately located at $\kappa=q_{m}d$, peaks of a Lorentzian shape in $\left|B_{\kappa}\right|^{2}$ are formed. In Fig.\ \ref{figApp} we demonstrate the structure of $\left|B_{\kappa}\right|^{2}$ by plotting it for the chain $n_{0}=10$, $r=5$.

\begin{figure} [!h]
 \begin{center}
 \includegraphics[width=0.45\columnwidth]{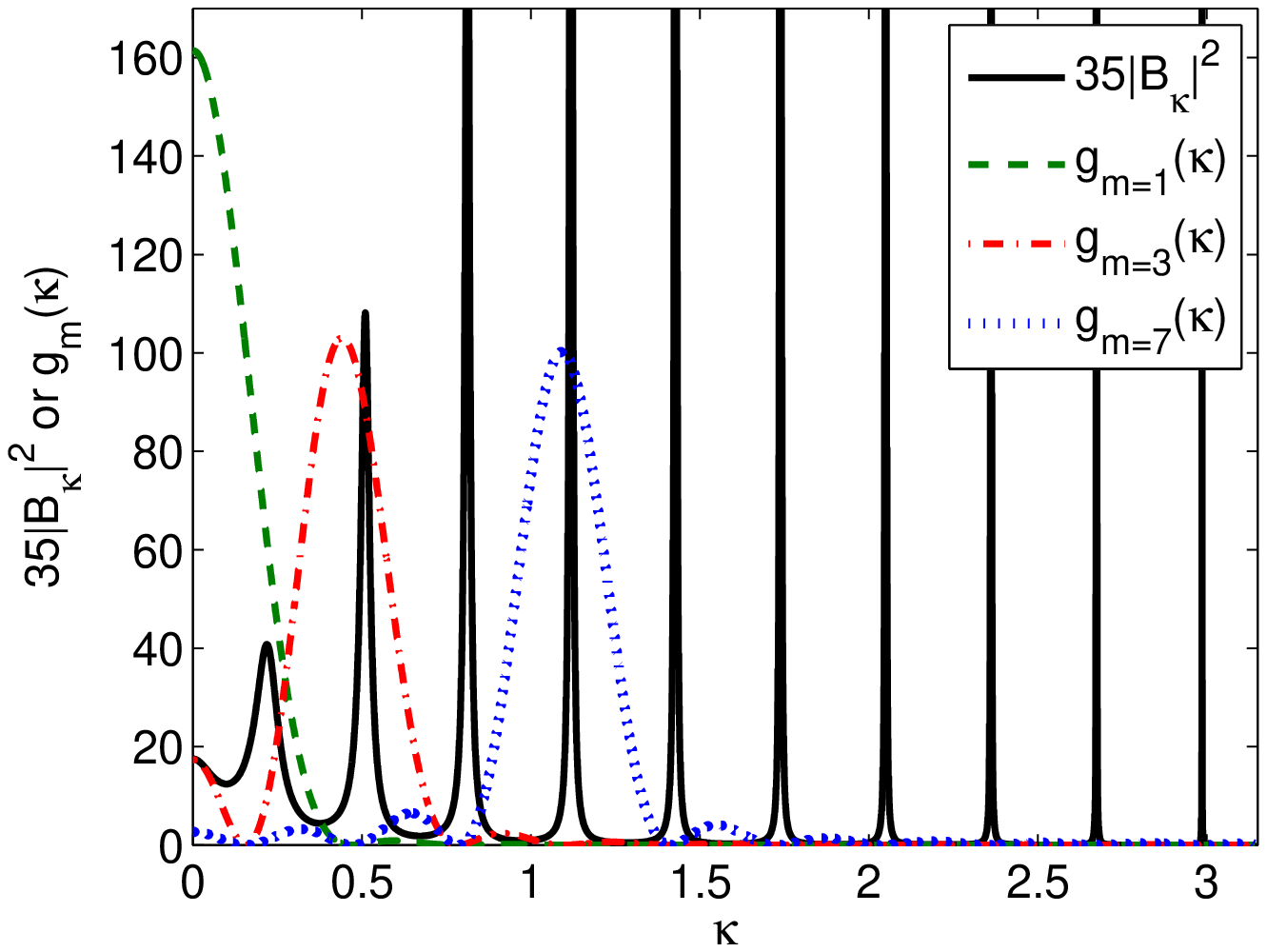}
 \includegraphics[width=0.45\columnwidth]{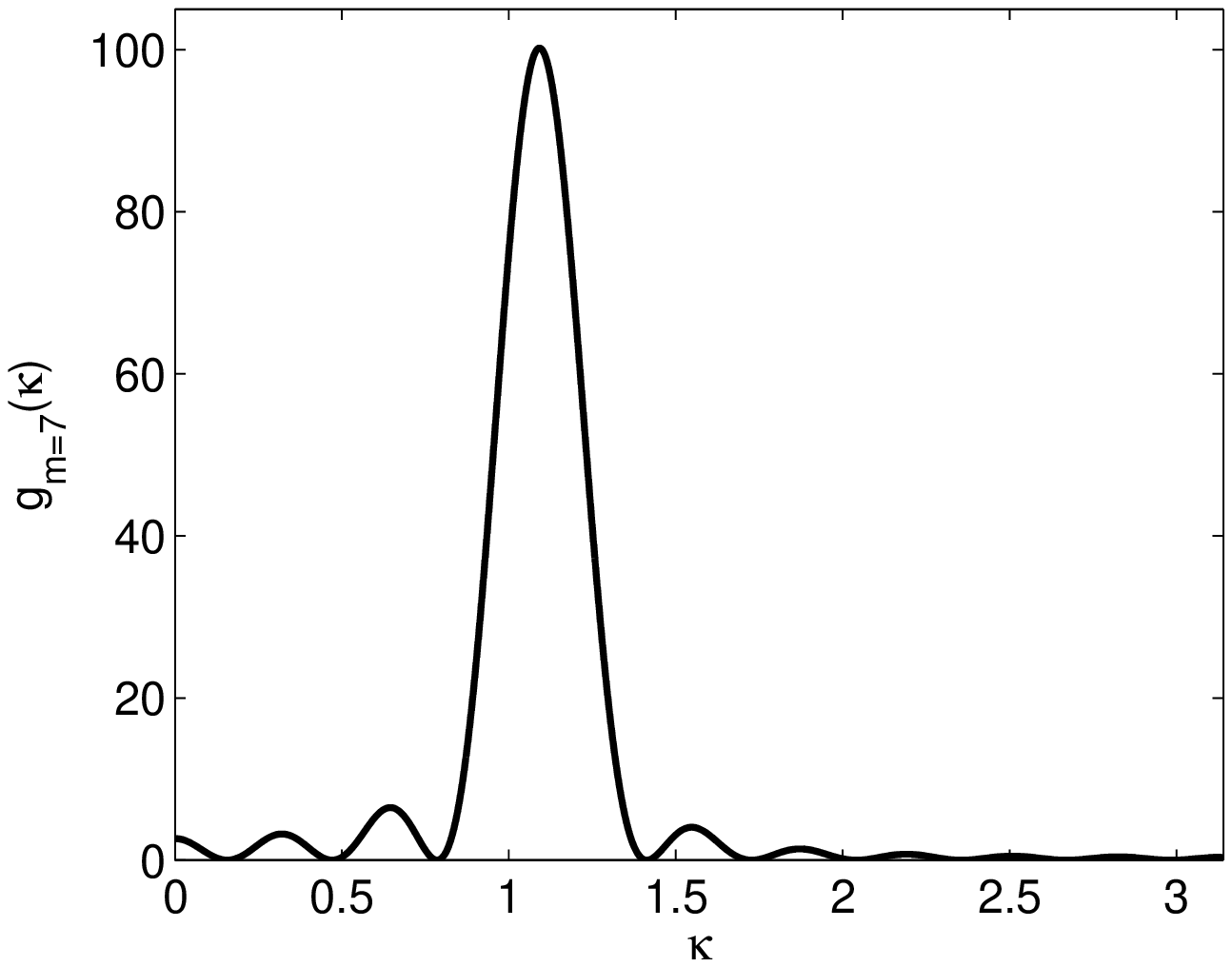}
 \caption{\label{figgpp}
  (Left, color online) The function $g_{m}\left(\kappa\right)$ of the modes $m=1$, $m=3$ and $m=7$ of a chain with $n_{0}=10$ and $r=5$, plotted together with $\left|B_{\kappa}\right|^{2}$ for symmetrical modes. The function $\left|B_{\kappa}\right|^{2}$ is multiplied by 35 in order to facilitate the comparison between the location of the main peak of $g_{m}\left(\kappa\right)$ and the location of the resonances $\left|B_{\kappa}\right|^{2}$. (Right) The function $g_{m}\left(\kappa\right)$ for $m=7$ shown on its own in order to highlight the structure of the secondary peaks.}
 \end{center}
\end{figure}

In addition, in order to obtain a nonvanishing value of the spectral function the function $g_{m}\left(\kappa\right)$ of Eq.\ \eqref{eq:g} cannot be too small. This function describes the overlap between the mode of the infinite chain with a wave number $\kappa$, and the eigenmode of the finite chain with a wave number $q_{m}d$. The general structure of $g_{m}\left(\kappa\right)$ consists of a main peak located approximately at $k_{0}=q_{m}$, due to constructive interference occurring when the the two modes are synchronized, and much lower secondary peaks. Half the distance between the two minima of $g(\kappa)$ closest to the main maximum is equal to $\pi/n_{0}$, which is also the distance between adjacent modes of equal parity. Thus, $g_{m}\left(\kappa\right)$ selects mainly the resonance in $\left|B_{\kappa}\right|^{2}$ that is located around $\kappa\simeq q_{m}d$. In Fig.\ \ref{figgpp} we demonstrate these features of $g_{m}\left(\kappa\right)$ for several modes of the chain with $n_{0}=10$ and $r=5$.

Therefore, the main contribution to the spectral function of a given mode is due to the resonance of $\left|B_{\kappa}\right|^{2}$ around $\kappa\simeq q_{m}d$, and the quality of the Lorentzian approximation is, to a large extent, determined by the amount of overlap with the adjacent resonances of $\left|B_{\kappa}\right|^{2}$. If the widths of the resonances are small compared to the distance between them then the contribution from the adjacent resonances is small and the spectral function of the mode is narrow and Lorentzian-like. Changes in parameters such as $m$, $n_{0}$, and $r$ can improve or worsen the quality of the Lorentzian approximation. In the present study we consider modes and chains for which the damping due to the environment is sufficiently small so the Lorentzian approximation is valid. Thus, the effect of the finite size of the chain is well pronounced and it influences the damping that results from the phonon-phonon interaction that we consider in the next section. The applicability of the Lorentzian approximation also allows us to describe the clamping-losses using a single constant---the quality factor $Q^{\mathrm{c}}_{m}$.

However, for the chains considered here the Lorentizan approximation fails for the lowest vibrational mode $m=1$, as the tail of the spectral function near $\Omega=0$ is higher than the peak at $\Omega^{*}_{q_{m=1}}$ and the width of the spectral function is comparable with $\Omega^{*}_{q_{m=1}}$. Therefore, we usually do not discuss the damping due to phonon-phonon interaction of phonons in the lowest vibrational mode interaction, which in any case is very small compared to the clamping-losses. The large tail at $\Omega=0$ occurs due to the ease in which long wavelength modes of the infinite chain penetrate into the finite chain, combined with the fact that the main peak of $g_{m=1}\left(\kappa\right)$ is located at $\kappa=0$ due to the almost rigid translation of the finite chain in the limit $\kappa\rightarrow0$.

\section{Anharmonic Interaction Within the Finite Chain}\label{sec:Anharmonic interaction within the finite chain}

\subsection{Anharmonic Hamiltonian}\label{subsec:The anharmonic Hamiltonian}
We consider anharmonic phonon-phonon interaction that is restricted to the finite atomic chain. We use the Green function of the previous section in order to incorporate the effect of the outer environment into the calculation of the damping. We take into account the lowest order nearest-neighbor anharmonic potential $U_{3}$ within the finite chain
\begin{align}\label{eq:U3}
U_{3}=&\frac{1}{6}\sum_{\alpha,\beta,\gamma}\sum_{n=-n_{0}}^{n_{0}-1}\phi_{\alpha\beta\gamma}\left[u^{\mathrm{fin}}_{\alpha}(n+1,t)
-u^{\mathrm{fin}}_{\alpha}(n,t)\right]\left[u^{\mathrm{fin}}_{\beta}(n+1,t)-u^{\mathrm{fin}}_{\beta}(n,t)\right]\nonumber\\
&\times\left[u^{\mathrm{fin}}_{\gamma}(n+1,t)-u^{\mathrm{fin}}_{\gamma}(n,t)\right]
\end{align}
where $\phi_{\alpha\beta\gamma}$ are the cubic anharmonic nearest-neighbor force constants, and
\begin{equation}\label{u time dep}
u^{\mathrm{fin}}_{\alpha}(n,t)=\sum_{m}\psi_{m,\alpha}\left(n\right)e^{i\omega_{m,\alpha}t}
\end{equation}
is the displacement field within the clamped finite chain, with $\psi_{m,\alpha}\left(n\right)$ given by Eq.\ \eqref{eq:finite modes}. As a result, we consider the effect of only three-phonon processes while higher order processes are neglected. The anharmonic interaction $U_{3}$ induces an additional damping of the vibrational modes of the finite chain in addition to the damping caused by clamping-losses. As for clamping-losses, the damping due to three-phonon processes is due to the fact that the modes of the finite chain are not eigenmodes of the full Hamiltonian of the system when $U_{3}$ is included in it. The externally added phonons again decay into a continuum, which, in the case of the phonon-phonon interaction, is formed by the phonons themselves.

We insert the quantized displacement field \eqref{eq:quantization finite chain} into Eq.\ \eqref{eq:U3} and obtain
\begin{align}
U_{3}=&\frac{1}{6}\sum_{\alpha,\beta,\gamma}\phi_{\alpha\beta\gamma}\sum_{m_{1},m_{2},m_{3}}
\left(\frac{\hbar}{2m_{\arm}}\right)^{\frac{3}{2}}\frac{1}{\sqrt{\omega_{m_{1},\alpha}\omega_{m_{2},\beta}\omega_{m_{3},\gamma}}}
A_{m_{1},\alpha}A_{m_{2},\beta}A_{m_{3},\gamma}\nonumber\\
&\times\sum_{n=-n_{0}}^{n_{0}-1}\left(\psi_{m_{1},\alpha}\left(n+1\right)-\psi_{m_{1},\alpha}\left(n\right)\right)
\left(\psi_{m_{2},\beta}\left(n+1\right)-\psi_{m_{2},\beta}\left(n\right)\right)\nonumber\\
&\times\left(\psi_{m_{3},\gamma}\left(n+1\right)-\psi_{m_{3},\gamma}\left(n\right)\right)\nonumber\\
=&\sum_{\alpha,\beta,\gamma}\sum_{m_{1},m_{2},m_{3}}
H_{\alpha\beta\gamma}^{m_{1}m_{2}m_{3}}\Delta\left(m_{1},m_{2},m_{3}\right)A_{m_{1},\alpha}A_{m_{2},\beta}A_{m_{3},\gamma},
\label{eq:U3 quantized}
\end{align}
where
\begin{equation}\label{eq:H}
H_{\alpha\beta\gamma}^{m_{1}m_{2}m_{3}}=\frac{2\phi_{\alpha\beta\gamma}}{3\sqrt{n_{0}}}\left(\frac{\hbar}{2m_{\arm}}\right)^{\frac{3}{2}}
\frac{\sqrt{\omega_{m_{1},\alpha}\omega_{m_{2},\beta}\omega_{m_{3},\gamma}}}{\omega_{0,\alpha}\omega_{0,\beta}\omega_{0,\gamma}},
\end{equation}
and the summation over the atomic sites (the index $n$) in Eq.\ \eqref{eq:U3 quantized} results in momentum conservation, which is expressed by the function
\begin{equation}\label{eq:Delta m}
\Delta\left(m_{1},m_{2},m_{3}\right)=
\left(\delta_{m_{1}-m_{2}+m_{3},0}+\delta_{m_{1}+m_{2}-m_{3},0}+\delta_{m_{1}-m_{2}-m_{3},0}+\delta_{m_{1}+m_{2}+m_{3},4n_{0}}\right),
\end{equation}
where the first three Kronecker $\delta$ functions reflect momentum conserving three-phonon processes, and the fourth represents an umklapp process.

We estimate the value of the cubic force constants $\phi_{\alpha\beta\gamma}$ using the quadratic force constants $\phi_{\alpha}$ and their relation to the Gruneisen parameters $\gamma_{\alpha}$ in a linear chain \cite{ashcroft,goldfarb}
\begin{equation}\label{eq:gruneisen}
\gamma_{\alpha}=\frac{d}{2}\frac{\phi_{x\alpha\alpha}}{\phi_{\alpha}}.
\end{equation}
In light of our simplified treatment and our rough estimate of the anharmonic force constants, we use a single averaged third order force constant when carrying out the actual numerical calculations.

\subsection{Damping due to phonon-phonon interaction}\label{subsec:Anharmonic lifetimes of the vibrational modes of the finite chain}

\begin{figure} [!h]
 \begin{center}
 \includegraphics[width=0.45\columnwidth]{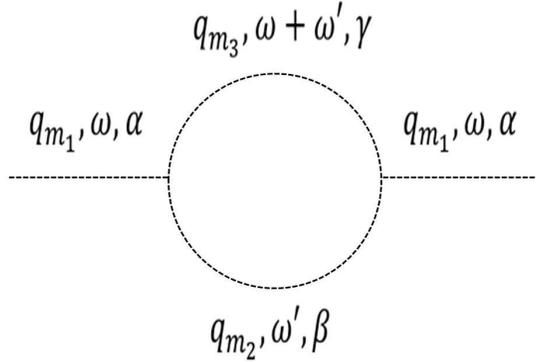}
 \caption{\label{figdiag}
 The lowest order correction to the self-energy of the phonons due to three-phonon processes. Each line in the diagram corresponds to a Green function $D^{0}$ that is defined in Eq.\ \eqref{eq:green phonon}, with momentum, frequency, and polarization as labeled in the diagram.}
 \end{center}
\end{figure}

We calculate the imaginary part of the self-energy (IPSE) of the phonons of the finite chain, using the Green function given in Eq.\ \eqref{eq:green phonon} which incorporates the effect of the coupling to the environment. The IPSE is calculated to the lowest order in the three-phonon processes,\cite{mahan} as shown in the diagram in Fig.\ \ref{figdiag}
\begin{align}
\Gamma^{\mathrm{ph}}_{m_{1},\alpha}\left(\omega\right)=&\mathrm{Im}\left[\lim_{\epsilon'\rightarrow 0}\lim_{\epsilon\rightarrow 0} -i\frac{9\hbar}{\pi}\sum_{\beta,\gamma}\sum_{m_{2},m_{3}}\Delta\left(m_{1},m_{2},m_{3}\right)\right.\nonumber\\
&\left.\times\int d\omega^{'} \left|H_{\alpha\beta\gamma}^{m_{1}m_{2}m_{3}}\right|^{2}D_{m_{2},\beta}^{0}\left(\omega^{'}\right)D_{m_{3},\gamma}^{0}\left(\omega+\omega^{'}\right)\right],
\label{eq:Gamma1}
\end{align}
where the infinitesimal factors $\epsilon$ and $\epsilon'$ refer to the Green functions $D_{m_{2},\beta}^{0}\left(\omega^{'}\right)$ and \newline $D_{m_{3},\gamma}^{0}\left(\omega+\omega^{'}\right)$, as given in Eq.\ \eqref{eq:green phonon}.

We characterize the damping of the vibrational modes by the value of the IPSE at $\omega_{m,\alpha}^{*}$. In using this approximation we neglect the shift of the frequencies of the modes due to the phonon-phonon interaction, and assume that the entire IPSE can be replaced by its value at the nonrenormalized frequency $\omega_{m,\alpha}^{*}$. We therefore assume that the spectral function of the phonons in the presence of the phonon-phonon interaction can still be approximated as a Lorentzian whose peak is located at $\omega=\omega_{m,\alpha}^{*}$. This assumption is commonly used in the calculation of the lifetime of phonons in bulk systems \cite{kokkedee2,maradudin,pathak} as well as in nanomechanical systems.\ \cite{martino} Thus, the contribution of the phonon-phonon interaction to the damping is given by
\begin{equation}\label{eq:Qm1 anh}
\left(Q^{\mathrm{ph}}_{m,\alpha}\right)^{-1}=\frac{2\Gamma^{\mathrm{ph}}_{m,\alpha}\left(\omega_{m,\alpha}^{*}\right)}{\hbar\omega_{m,\alpha}^{*}},
\end{equation}
where $\omega_{m,\alpha}^{*}$ corresponds to the frequency that maximizes the spectral function \eqref{eq:spectral fun final}. If the Lorentzian approximation fails, then one needs to evaluate the spectral function in order to describe the damping of the vibrational mode, for which one requires both the frequency dependent real and imaginary parts of the self-energy. We note that if only the damping due to clamping-losses is taken into account then the condition for operating within the Landau-Rumer regime holds, in the chains considered here, for the decay processes which significantly contribute to $\qanh$.

We insert the expression for the Green function of the phonons given in Eq.\ \eqref{eq:green phonon} into Eq.\ \eqref{eq:Gamma1} and perform the integral over $\omega^{'}$ by means of contour integration, where only terms with one pole above and one pole below the real axis contribute to the final result. We change the sums over $k$ and $k'$ into integrals, take the limits $\epsilon,\epsilon'\rightarrow 0$ and then the imaginary part of the result, obtaining
\begin{align}
\Gamma^{\mathrm{ph}}_{m_{1},\alpha}\left(\omega\right)=&\frac{9d^{2}}{2\pi}\sum_{\beta,\gamma}\sum_{m_{2},m_{3}}\Delta\left(m_{1},m_{2},m_{3}\right)
\left|H_{\alpha\beta\gamma}^{m_{1}m_{2}m_{3}}\right|^{2}\nonumber\\
&\times\int^{\frac{\pi}{d}}_{-\frac{\pi}{d}}\int^{\frac{\pi}{d}}_{-\frac{\pi}{d}}dkdk' \left|\left\langle 1_{k,\beta}\right|A_{m_{2},\beta}\left|0\right\rangle\right|^{2}\left|\left\langle 1_{k',\gamma}\right|A_{m_{3},\gamma}\left|0\right\rangle\right|^{2}\nonumber\\
&\times\left[\delta\left(\omega-\omega_{k,\beta}-\omega_{k',\gamma}\right)-\delta\left(\omega+\omega_{k,\beta}+\omega_{k',\gamma}\right)\right].\label{eq:Gamma2}
\end{align}
Since we are interested in positive values of $\omega$, only the first $\delta$ function contributes to the IPSE.

Using the dispersion relation \eqref{eq:dispersion relation} of the modes of the infinite chain we obtain
\begin{equation}\label{eq:dos}
\frac{dk}{d\omega_{k,\alpha}}=\frac{2}{d}\frac{1}{\sqrt{\omega_{0,\alpha}^{2}-\omega_{k,\alpha}^{2}}},
\end{equation}
which is used to change the integral over the wave numbers $k$ and $k'$ in Eq.\ \eqref{eq:Gamma2} into integrals over the corresponding frequencies $\omega_{k}$ and $\omega_{k'}$. We perform the integral over $\omega_{k}$ and obtain the final expression for the IPSE
\begin{align}
\Gamma^{\mathrm{ph}}_{m_{1},\alpha}\left(\omega\right)=&\frac{72}{\pi}\sum_{\beta,\gamma}\sum_{m_{2},m_{3}}\Delta\left(m_{1},m_{2},m_{3}\right)
\left|H_{\alpha\beta\gamma}^{m_{1}m_{2}m_{3}}\right|^{2}\nonumber\\
&\times\int^{\omega_{k,max}}_{\omega_{k,min}}d\omega_{k}
\frac{\left|\left\langle 1_{\omega-\omega_{k},\beta}\right|A_{m_{2},\beta}\left|0\right\rangle\right|^{2}}{\sqrt{\omega_{0,\beta}^{2}-\left(\omega-\omega_{k}\right)^{2}}}
\frac{\left|\left\langle 1_{\omega_{k},\gamma}\right|A_{m_{3},\gamma}\left|0\right\rangle\right|^{2}}{\sqrt{\omega_{0,\gamma}^{2}-\omega_{k}^{2}}}
,\label{eq:Gamma final}
\end{align}
where $\omega_{k,max}=\mathrm{min}\left[\omega_{0,\gamma},\omega\right]$ and $\omega_{k,min}=\mathrm{max}\left[0,\omega-\omega_{0,\beta}\right]$, and we changed the integration variable in Eq.\ \eqref{eq:Gamma final} from $\omega_{k'}$ to $\omega_{k}$. The limits of integration $\omega_{k,min}$ and $\omega_{k,max}$ ensure that the arguments of the square-roots in Eq.\ \eqref{eq:Gamma final} are positive.

The expression in Eq.\ \eqref{eq:Gamma final} is a sum over convolutions of pairs of spectral functions of the type appearing in Eq. \eqref{eq:spectral fun final}. Each convolution represents a decay of the original phonon of a wave number $q_{m_{1}}$, frequency $\omega$, and polarization $\alpha$ into a phonon  with a spectral function that is centered around $\omega_{m_{3},\gamma}^{*}$ and a second phonon with a spectral function that is centered around the shifted frequency $\omega-\omega_{m_{2},\beta}^{*}$. A maximal overlap between the spectral functions occurs if their peaks are at the same position, namely $\omega_{m_{3},\gamma}^{*}=\omega-\omega_{m_{2},\beta}^{*}$. In other words, in order to obtain a significant contribution from one of the terms in Eq.\ \eqref{eq:Gamma final} it is necessary for the decay process to approximately conserve energy. Damping is especially large if the resulting phonons in a decay process that approximately conserves energy possess spectral functions with similar shapes, which means that their mode numbers cannot be too much separated one from the other.

\subsection{Phonon-phonon interaction -- results and discussion}\label{subec:Results unharmonic}

We demonstrate the effect of the approximately discrete nature of the phonon states on the damping due to phonon-phonon interaction by evaluating the damping as a function of three parameters---the mode number $m$ of the excited mode, the ratio between the transverse and longitudinal speeds of sound $R=\sqrt{\phi_{t}/\phi_{l}}=c_{t}/c_{l}$, where $0<R<1$, and the number of atoms $2n_{0}-1$ in the finite chain. Varying each parameter, we observe peaks or resonances in the damping of the excited mode whenever an energy-conserving decay process is possible. The inability to fulfill energy conservation for most of the modes, resulting in low $\qanh$, is due to the discrete-like nature of the finite-chain modes, which are still resolved if the coupling to the infinite chain is sufficiently weak. Thus, the resonances in $\qanh$ are a reflection of the effect of the finite size of the chain on its vibrational modes. We found similar peaks and resonances in the damping, although of a somewhat different origin, in the study of electron-phonon damping in metallic nanomechanical beams.\ \cite{lindenfeld2} Resonances in the dissipation were also observed experimentally, but in a different system and regime of operation, by tuning the frequency of the fundamental flexural mode of a thin SiN membrane.\ \cite{jockel1}

Since the transverse branches lie below the longitudinal branch, it is impossible for any transverse phonon in a linear chain [i.e.\ with a dispersion relation that is given by Eq.\ \eqref{eq:dispersion relation}] to decay into two phonons while simultaneously conserving momentum and energy.\ \cite{ashcroft,goldfarb} Thus, the damping of transverse modes does not exhibit sharp peaks. On the other hand, the energy conservation equation, $\left|\sin{\frac{\kappa_{1}}{2}}\right|=\left|\sin{\frac{\kappa_{2}}{2}}\right|+R\left|\sin{\frac{\kappa_{3}}{2}}\right|$, for a decay process of the type $l\rightarrow l+t$, can be solved while simultaneously conserving momentum, if $\kappa_{2,3}$ are continuous wave numbers.\ \cite{ashcroft,goldfarb} The same is true also for decay processes of the type $l\rightarrow t+t$. This opens the possibility for an almost exact energy conservation in the decay of some sufficiently high-frequency longitudinal modes, which can lead to larger damping. Note that from now on we use $l$ to denote the longitudinal polarization while the letter $t$ is used to denote either the $y$ or $z$ transverse polarizations.

In order to obtain numerical results for the contribution of phonon-phonon interaction to the damping we use average material properties of bulk silicon, given in Table \ref{properties_silicon}. We again point out that we use these parameters in order to obtain rough qualitative estimates, and they by no means reflect the exact properties of an actual silicon atomic chain.

\begin{table}[!h]
 \begin{center}
 \begin{tabular}{|l|l|}
 \hline\hline
 $c_{l}$    & $\mathrm{8.5}\times10^{3}\frac{m}{sec}$\\
 $c_{t}$    & $\mathrm{5.9}\times10^{3}\frac{m}{sec}$\\
 $d$        & 0.5nm\\
 $\gamma_{l}$    & 1.1\\
 $\gamma_{t}$ & 0.32\\
\hline\hline
 \end{tabular}
 \end{center}
\caption{\label{properties_silicon}
 Averaged material properties of bulk silicon used for calculating the results in section \ref{subec:Results unharmonic}.\ \cite{dargys}}
\end{table}

\begin{figure} [!h]
 \begin{center}
 \includegraphics{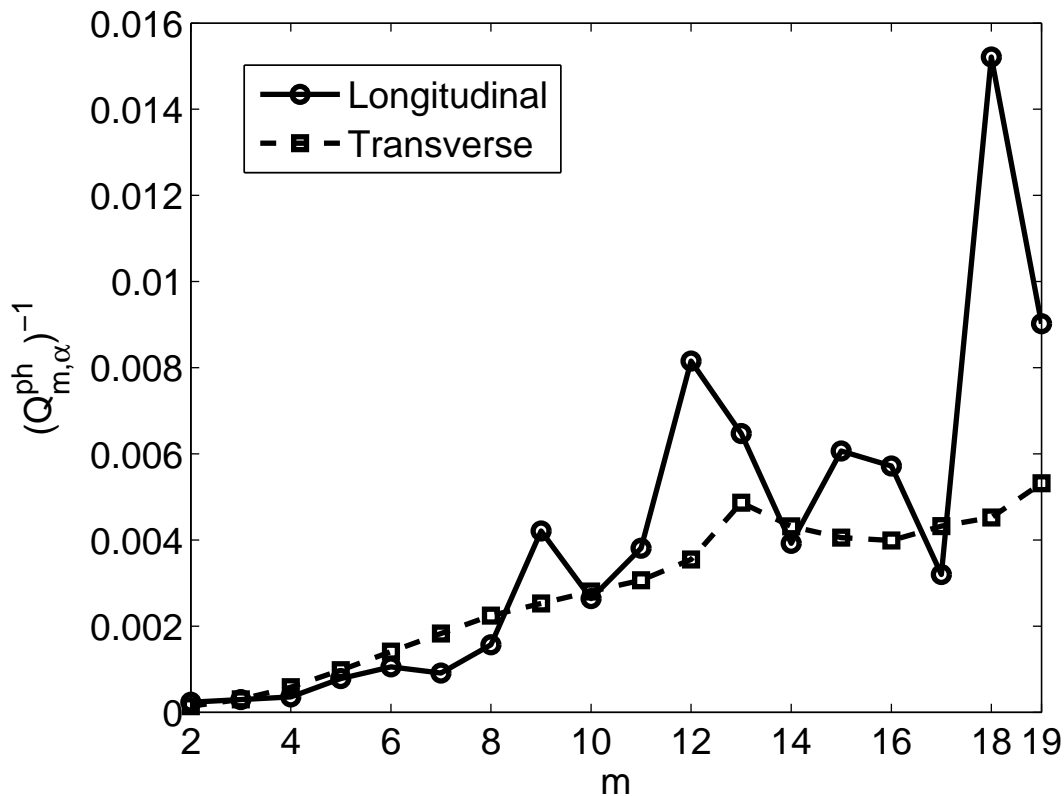}
 \caption{\label{fig3pp}
 Damping due to phonon-phonon interaction of the longitudinal and transverse modes of a chain with $n_{0}=10$ (19 atoms) and $r=5$.}
 \end{center}
\end{figure}

\begin{figure} [!h]
 \begin{center}
 \includegraphics{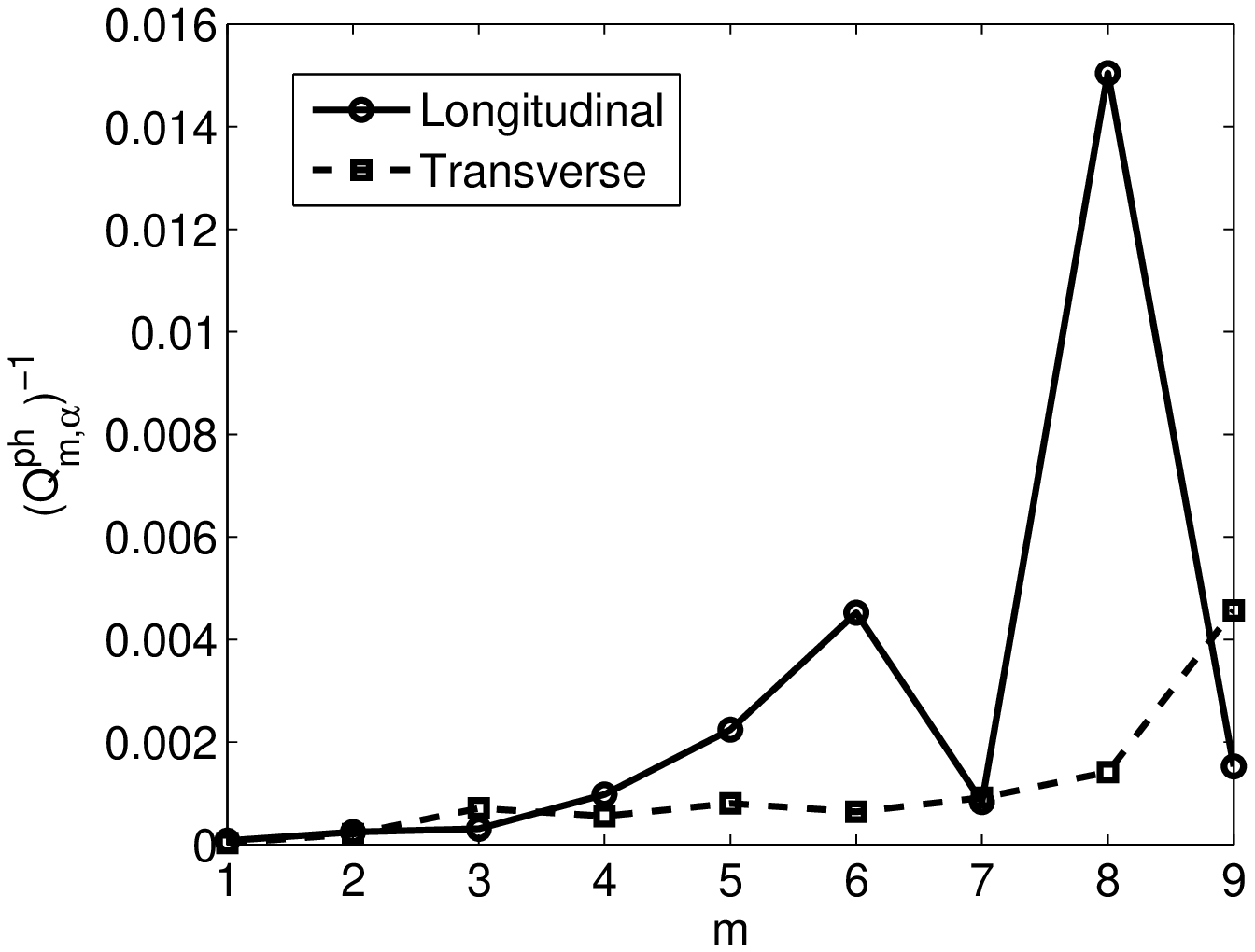}
 \caption{\label{fig5pp}
 Damping due to phonon-phonon interaction of the longitudinal and transverse modes of a chains with $n_{0}=5$ and $M_{\arm}/m_{\arm}=15$.}
 \end{center}
\end{figure}

We consider once again a chain with $n_{0}=10$ and $r=5$ and present $\qanh$ of the longitudinal and transverse modes of the chain in Fig.\ \ref{fig3pp}. The sharp peaks in the dissipation for the longitudinal modes with $m=9,12,15,$ and $18$ are due to decay processes that approximately conserve energy. For example, for $m=18$ the damping of the excited phonon is dominated by the process $[m=18,\alpha=l]\rightarrow [m=13,\beta=l]+[m=5,\gamma=t]$ and the umklapp process $[m=18,\alpha=l]\rightarrow [m=16,\beta=t]+[m=6,\gamma=t]$, both of which approximately conserve energy. A second example for the same behavior is given in Fig.\ \ref{fig5pp} for a shorter chain with a larger $r$ ratio (less clamping-losses).

Low-frequency phonons cannot decay while fulfilling the energy conservation requirement. Furthermore, most decay processes of such phonons do not contribute much to $\qanh$ since they involve a decay into two phonons of which at least one is more energetic than the energy of the excited mode. As the energy of the excited mode increases with the increase in $m$, the number of processes that involve a decay into two less energetic phonons increases. This leads to the general trend of a gradual increase in $\qanh$ with the increase in $m$.

\begin{figure} [!h]
 \begin{center}
 \includegraphics{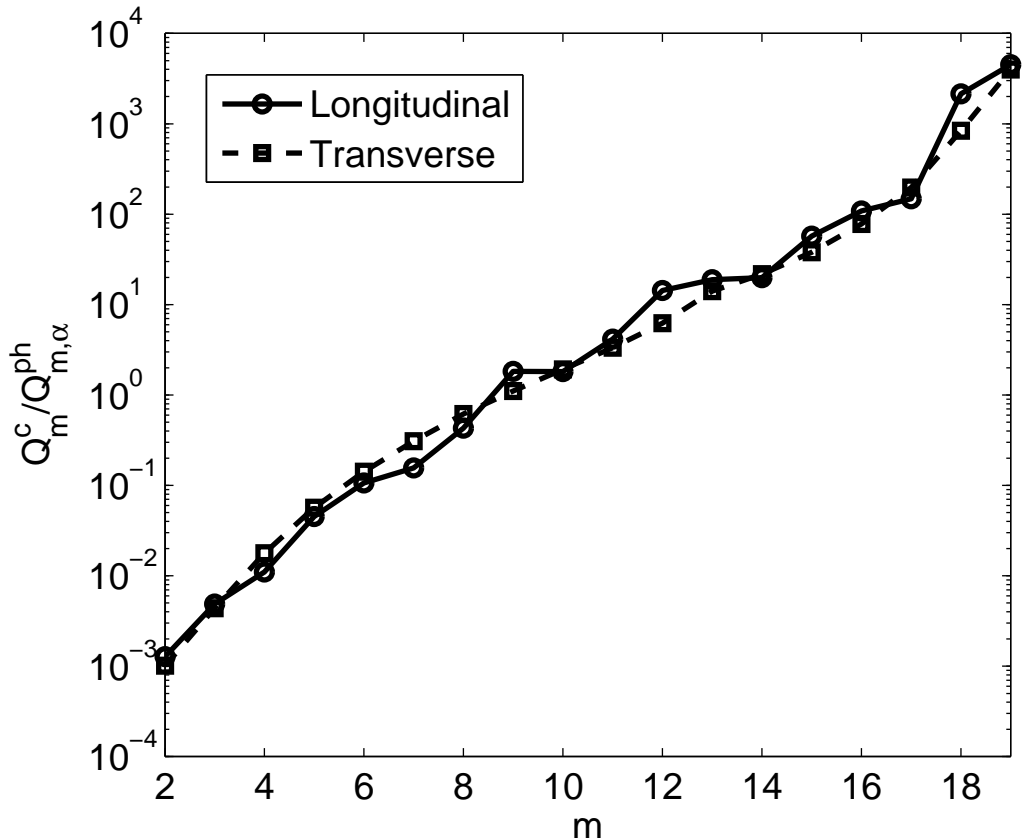}
 \caption{\label{fig4pp}
 The ratio between damping due to phonon-phonon interaction and damping due to clamping-losses for longitudinal and transverse modes of the chain considered in Fig.\ \ref{fig3pp}.}
 \end{center}
\end{figure}

In Fig.\ \ref{fig4pp} we compare the contribution of clamping-losses to the damping of a given mode to the damping caused by phonon-phonon interaction. As the mode number increases, the damping due to clamping-losses decreases, while the damping due to phonon-phonon interaction increases. This usually means that for a certain mode number the damping due to phonon-phonon interaction becomes more important than the clamping-losses, even in the absence of a decay process that approximately conserves energy. However, the decrease in clamping-losses with the increase in the mode number may well be related to the specific type of outer environment we chose in our model, and its coupling to the finite chain. Thus, the importance of the damping of a specific mode due to phonon-phonon interaction compared to other dissipation mechanisms, and specifically clamping-losses, may strongly depend on the system under consideration.

Damping of a given mode can be controlled by changing the ratio $R$. Changing $R$ might be achieved by stretching the chain, which is likely to induce a larger modification in $c_{l}$ than in $c_{t}$. Thus, in the decay processes of the type $l\rightarrow t+t$ or $l\rightarrow l+t$, the change in $R$ can change the sum of the energies of the resulting phonons compared to the energy of the excited mode. This can improve or worsen the energy conservation of a decay process and thus increase or decrease its contribution to $\qanh$. Such manipulation can affect the energy conservation in a decay process only if at least one of the product phonons is of a different polarization from that of the excited mode. Except for the separate influence of $c_{t}$ and $c_{l}$ on the anharmonic force constants through Eq.\ \eqref{eq:gruneisen}, only the ratio between the transverse and longitudinal speeds of sound affects $\qanh$.

\begin{figure} [!h]
 \begin{center}
 \includegraphics{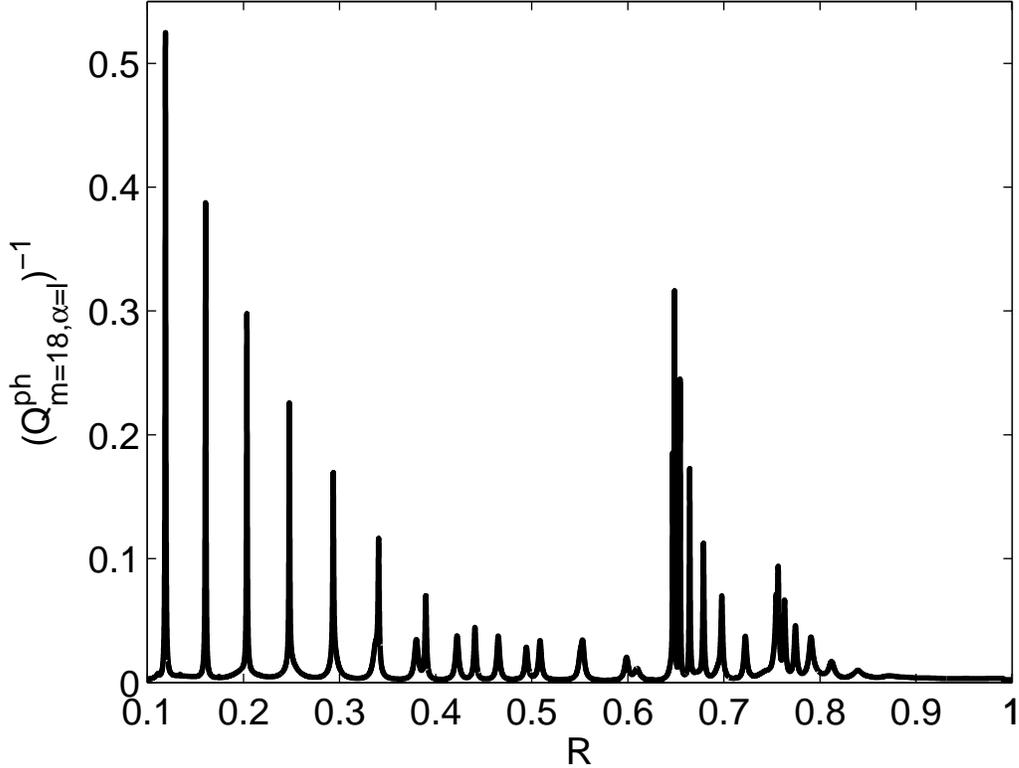}
 \caption{\label{fig7Rpp}
 The damping due to phonon-phonon interaction of the $18^{\mathrm{th}}$ longitudinal mode of the chain considered in Fig.\ \ref{fig3pp}, is shown as a function of the ratio $R=c_{t}/c_{l}$. The material parameters given in Table \ref{properties_silicon} result in $R\approx 0.69$. The two peaks located between $R\approx0.05$ and $R\approx0.1$ are not shown.}
 \end{center}
\end{figure}

In  Fig.\ \ref{fig7Rpp} we demonstrate the resonance structure in the damping of the $18^{\mathrm{th}}$ longitudinal mode that is obtained when $R$ is varied. Each of the peaks in Fig.\ \ref{fig7Rpp} corresponds to one or two decay processes that become almost energy-conserving and therefore dominate the damping. For example, the approximately equally-spaced peaks that we find for small values of $R$ correspond to the umklapp processes $\left[m=18,\alpha=l\right]\rightarrow\left[m=15,\beta=t\right]+\left[m=7,\gamma=t\right]$,  $\left[m=18,\alpha=l\right]\rightarrow\left[m=14,\beta=t\right]+\left[m=8,\gamma=t\right]$, and so on.

The frequency of the longitudinal mode needs to be sufficiently large compared to the frequency of the modes of the transverse branch in order for an energy-conserving decay process to be possible. Thus, as $R$ increases above a certain value ($R\gtrsim0.9$ in Fig.\ \ref{fig7Rpp}) energy conservation becomes impossible and no peaks in the dissipation are observed. On the other hand, if $R$ is too small ($R\lesssim0.05$ for the mode considered in Fig.\ \ref{fig7Rpp}) the longitudinal mode is too energetic compared to the transverse modes and again energy conservation is impossible.

\begin{figure} [!h]
 \begin{center}
 \includegraphics[width=0.45\columnwidth]{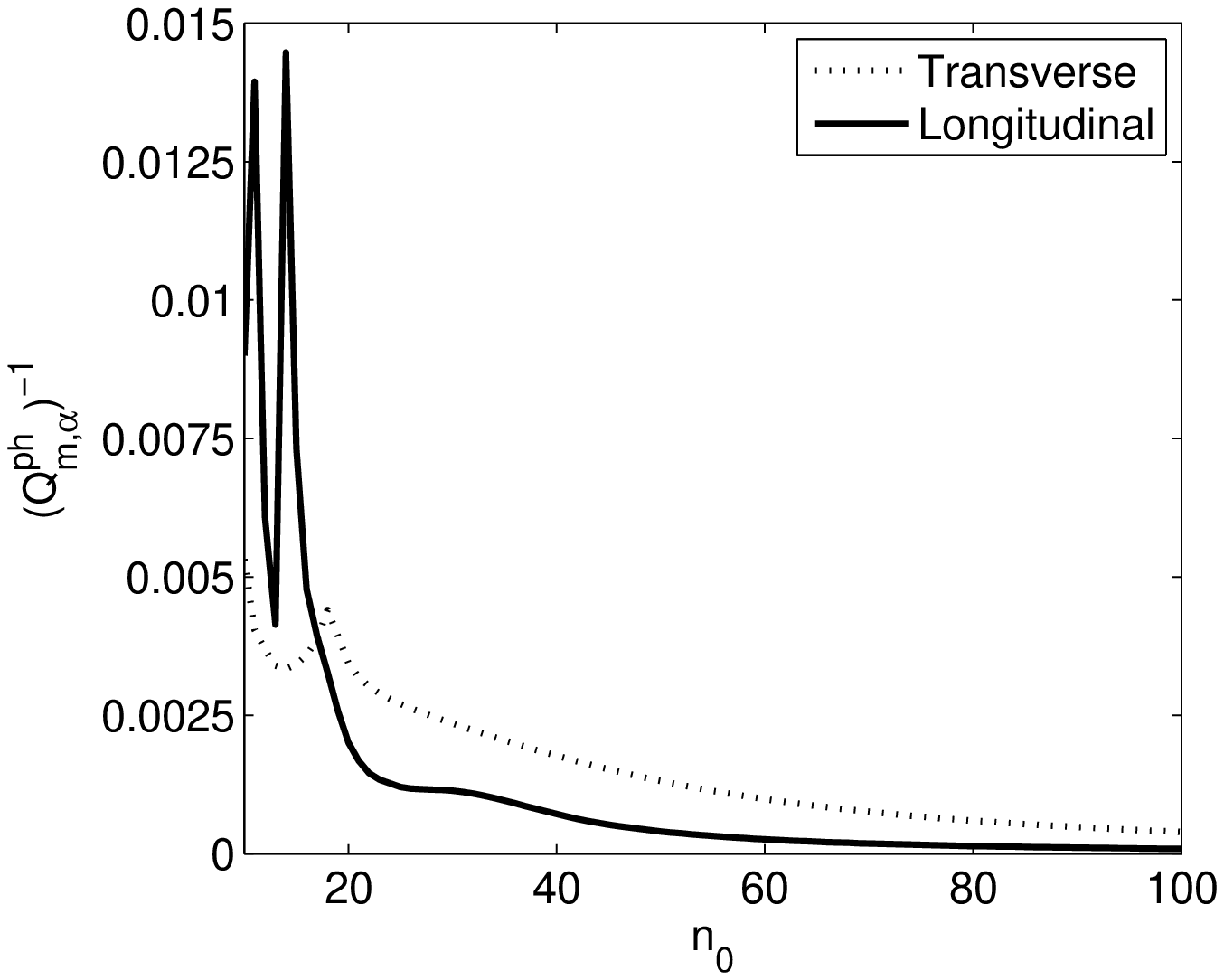}
 \includegraphics[width=0.45\columnwidth]{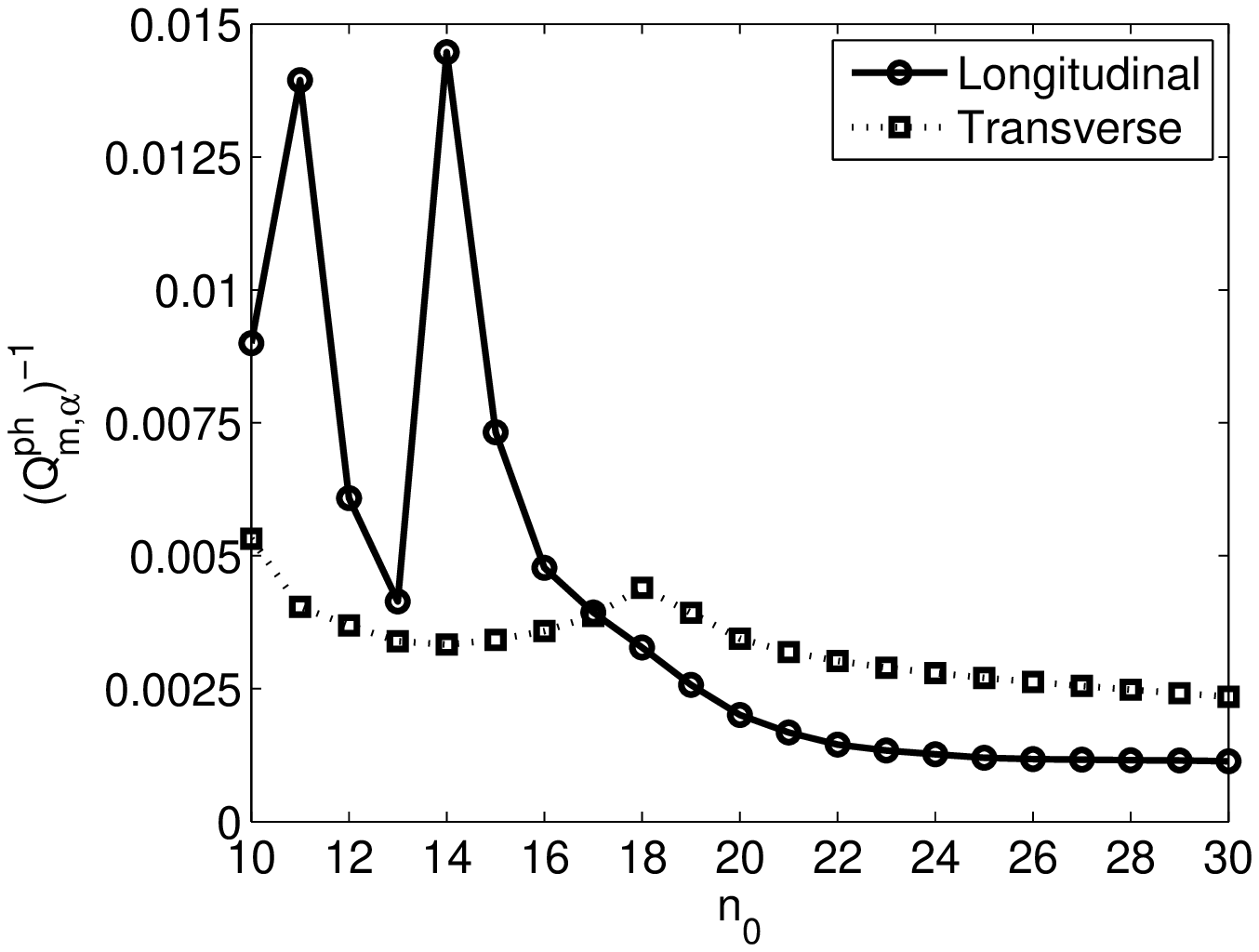}
 \caption{\label{fig6pp}
 (Left) Damping due to phonon-phonon interaction of the $19^{\mathrm{th}}$ longitudinal and transverse modes in chains with $r=5$, when the number of the atoms of the chain is varied from the shortest chain that supports the mode ($n_{0}=10$) and up to $n_{0}=100$. (Right) Zoom-in on chains with $n_{0}=10-30$.}
 \end{center}
\end{figure}

Lastly, we also consider $\qanh$ as a function of the number of atoms in the chain. The inverse quality factor of the $19^{\mathrm{th}}$ longitudinal and transverse modes as a function of $n_{0}$ is shown in Fig.\ \ref{fig6pp}. The change in the length of the chain shifts the frequencies of the phonon in the excited mode and of the phonons into which it decays, and therefore it changes the condition for energy conservation in existing decay processes. This might turn a process that does not conserve energy into one that approximately does, thus causing a large increase in the damping. This is demonstrated by the sharp increase in the damping in the chain with $n_{0}=11$, where the umklapp decay process $\left[m=19,\alpha=l\right]\rightarrow \left[m=19,\beta=t\right]+\left[m=6,\gamma=t\right]$ approximately conserves energy. In addition, by increasing the length of the chain, we introduce additional decay processes that might approximately conserve energy. This is the case for the peak in $\qanh$ at $n_{0}=14$, where energy conservation is improved for the process $\left[m=19,\alpha=l\right]\rightarrow \left[m=24,\beta=t\right]+\left[m=5,\gamma=t\right]$, which is absent up to $n_{0}=13$.

As $n_{0}$ is increased beyond a certain value, energy conservation in the additional processes becomes impossible since the frequency of the excited mode becomes lower than the maximal frequency of the transverse branch. Thus, peaks in the damping are not observed for sufficiently long chains. Also, the anharmonic coupling coefficient $H_{\alpha\beta\gamma}^{m_{1}m_{2}m_{3}}$ given in Eq.\ \eqref{eq:H} tends to zero, as $n_{0}$ increases for a given decay process of an excited mode with a fixed mode number. This leads to the vanishing of $\qanh$ at large $n_{0}$ that can be seen in Fig.\ \ref{fig6pp}.

\section{Summary and Conclusions}\label{sec:Conclusions pp}

We have studied damping due to phonon-phonon interaction within a finite atomic chain coupled by two heavy masses to two semi-infinite chains that form a continuum of external environment states. In addition to the simplicity of the model we used, our calculation also involved several additional restrictions and approximations: (1) The entire calculation was restricted to zero temperature; (2) only three-phonon processes were taken into account (3) the system was assumed to operate within the Landau-Rumer regime ($\omega\tau>1$); (4) the Lorentzian line-shape approximation for the spectral function was assumed to hold in the presence of the phonon-phonon interaction; (5) numerical results were obtained for the damping due to phonon-phonon interaction using averaged material parameters of bulk silicon.

Sharp peaks are formed in the damping caused by the phonon-phonon interaction as different parameters that characterize the excited mode or the finite atomic chain are varied. This behavior reflects the originally discrete nature of the vibrational modes, which is approximately maintained in the presence of a sufficiently weak coupling to the outer environment. The peaks in the damping appear when at least one of the decay processes of the phonon in the excited mode approximately conserves energy. Since we considered a longitudinal branch that lies above the transverse one, such decay processes are possible only for some sufficiently high-frequency longitudinal phonons. Nevertheless, even though the damping of most modes is governed by processes that do not conserve energy, their cumulative contribution can still result in non-negligible damping. Within the framework of our simplified model this damping can exceed the one obtained from clamping-losses.

The inclusion of higher-order phonon processes might result in the appearance of decay processes that approximately conserve energy, which are absent if only three-phonon processes are taken into account. This might lead to less pronounced peaks in the dissipation. Nevertheless, we expect that the basic structure of the damping will be maintained even if higher-order phonon processes are present since it reflects the approximately discrete nature of the spectral function of the phonons of the finite chain.

Our simple model can be used to study analytically, at least qualitatively, to what extent the damping due to phonon-phonon interaction within a finite system is sensitive to the details of the coupling to the outer environment, by considering different types of coupling and clamps. The dependence on temperature (as long as the system remains within the Landau-Rumer regime) can also be studied by repeating our calculation using the finite-temperature Green function formalism.

We expect that the qualitative features of the damping due to phonon-phonon interaction that we demonstrated in this study are quite general as long as the spectral function of the phonons is approximately discrete. Thus, we expect the sensitivity of the damping to changes in parameters to be present in other types of systems so long as those systems are sufficiently small, are only weakly coupled to an outer environment, with other internal dissipation mechanisms (such as electron-phonon interaction) being unimportant, and at a sufficiently low temperature. Such systems may include ultrasmall dielectric nanoparticles and sufficiently short non-metallic nanotubes or nanowires.

\bibliography{bibatomicchain}

\end{document}